\documentclass[twocolumn,prl,showpacs]{revtex4}

\usepackage{graphicx}
\usepackage{dcolumn}
\usepackage{bm}
\usepackage{amsmath}
\begin{document}
\preprint{Odd-frequency proximity}
\title{Cooper Pair Shape in Normal-metal/Superconductor Junctions}
\author{ Yukio Tanaka$^{1,2}$ Yasuhiro Asano$^{3}$, and Alexander A. Golubov$^{4}$ }
\affiliation{$^1$Department of Applied Physics, Nagoya University, Nagoya, 464-8603,
Japan \\
$^2$ CREST Japan Science and Technology Cooperation (JST) 464-8603 Japan \\
$^3$ Department of Applied Physics, Hokkaido University, Sapporo 060-8628,
Japan \\
$^4$ Faculty of Science and Technology, University of Twente, Enschede,
7500AE, The Netherlands}
\date{\today}

\begin{abstract}
In $s$-wave superconductors the Cooper pair wave function is
isotropic in momentum space. This property may also be expected for
Cooper pairs entering a normal metal from a superconductor due to
the proximity effect. We show, however, that such a deduction is
incorrect and the pairing function in a normal metal is surprisingly
anisotropic because of quasiparticle interference. We calculate
angle resolved quasiparticle density of states in NS bilayers which
reflects such anisotropic shape of the pairing function. We also
propose a magneto-tunneling spectroscopy experiment which could
confirm our predictions.
\end{abstract}

\pacs{74.45.+c, 74.50.+r, 74.20.Rp}
\maketitle

%$^4$ National Institute of Advanced Industrial Science and Technology
%(AIST), Tsukuba, 305-8568, Japan }

%--- title ---

%--- author ---

%
%
%--- address ---

%
%-----------------------------------------------------------
%-----------------------------------------------------------
It is well known that Cooper pairs consisting of two electrons are
characterized by electric charge $2e$, macroscopic phase, internal
spin, and by time and orbital structures~\cite{deGennes}. The charge
$2e$ manifests itself in various experiments, like Shapiro steps,
flux quantization and excess current due to the Andreev reflection.
The macroscopic phase generates the Josephson
current~\cite{deGennes}. The internal spin structure is classified
into spin-triplet and spin-singlet states which can be identified by
nuclear magnetic resonance. Further, based on a symmetry with
respect to the internal time, superconducting state can belong to
the even-frequency or the odd-frequency symmetry
class~\cite{Berezinskii}. The orbital degree of freedom is described
by an angular momentum quantum number $l$. In $s$-wave
superconductors with $l=0$, the Cooper pair wave function is
spherically symmetric on the Fermi surface, i.e. an angular
structure of a Cooper pair is isotropic in momentum space. The shape
of Cooper pairs in $p$-wave ($l=1$) and $d$-wave ($l=2$)
superconductors is characterized by two-fold and four-fold
symmetries, respectively~\cite{Sigrist}. The well established
properties listed above hold in bulk superconductors. The presence
of perturbations like spin-flip or interface scattering may change
the symmetry of Cooper pairs. For instance, unusual odd-frequency
property of Cooper pairs in proximity structures was predicted in
recent studies~\cite{Bergeret,Golubov2007}. The shape of Cooper pair
wave function in non-uniform systems like superconducting junctions
is not necessarily the same as that in the bulk state. Despite the
extensive study of the proximity effect during several past decades,
rather little attention has been paid to the problem of Cooper pair
shape in non-uniform superconducting systems \cite{Belzig,GK}. This
issue is quite important in view of current interest to the physics
of superconducting nanostructures.
\par

%Besides these background, recently it has been clarified
%that odd-frequency Cooper pair \cite{Bergeret} is
%induced near the interface and the normal metal (N)
%in normal metal / superconductor (N/S) junctions
%due to the spatial inhomogeneity of the pair potential \cite{Golubov2007,Ueda}.
%It has been shown that
%the amplitude of the odd-frequency component
%dominates at energies when the local density of state (LDOS)
%in the N/S junctions has subgap peaks due to the
%McMillan-Rowell oscillations \cite{Rowell}.
%
%It is actually interesting problem to clarify the
%shape of both the odd-frequency pair amplitude
%and even-frequency one.
%At the same time, to propose a new experimental setup to detect the
%change of the shape of the Cooper pair in the non-uniform
%superconducting system is important.
%It will serve  for the
%the progress of the physics of nanostructured superconducting
%systems near future.
%\par
The aim of the present Letter is to clarify the consequences of
breakdown of translational symmetry in superconductors on the Cooper
pair shape. For this purpose, we study the proximity effect in quasi
two-dimensional normal metal / superconductor (N/S) junctions by
solving the Eilenberger equation, treating self-consistently the
spatial variation of the superconducting pair potential. We analyze
the pairing function and the local density of states (LDOS) in N/S
junctions with spin-singlet $s$-wave and spin-triplet $p_{x}$-wave
superconductors.
%%%%%%%%%%%%%%%%%%%%%%%%%%%%%%%%%%%%%%%%%%%%%%%%%%%%%%%%
%We focus on the spatial dependence of the angular dependence of
%pairing function.
%%%%%%%%%%%%%%%%%%%%%%%%%%%%%%%%%%%%%%%%%%%%%%%%%%%%%%%%%%%%
Surprisingly, the pairing function in a normal
metal turns out to be strongly anisotropic even in junctions with
$s$-wave superconductors.
%
%In $p_{x}$-wave case, an amplitude of the pairing function of
%odd-frequency component exceeds that of even-frequency one and
%also shows the anisotropic behavior.
%Such anisotropic shape of a Cooper pair drastically modifies LDOS.
%
%In low energy, the shape of odd-frequency pair amplitude and
%that of angle resolved LDOS becomes similar.
%
To detect the complex Cooper pair shape, we propose to use scanning
tunneling spectroscopy in rotating magnetic field. We show that the
calculated tunneling conductance exhibits complex patterns even in
the $s$-wave case.
%
%The above results will contribute to physics of the
%nanostructured superconducting systems.
\par

\begin{figure}[tb]
\begin{center}
\scalebox{0.8}{
\includegraphics[width=8cm,clip]{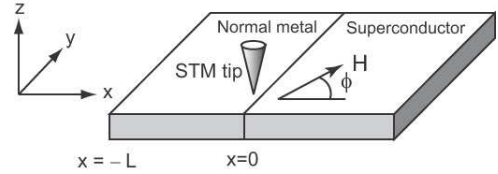}
}
\end{center}
\caption{A schematic illustration of a N/S junction.
}
\label{fig:1}
\end{figure}

Let us consider a quasi-two dimensional N/S junction as shown in
Fig.~\ref{fig:1} which is the simplest example of non-uniform
superconducting system, where the S region is semi-infinite and the
normal metal has finite length $L$.
%We assume that the effective mass along the $z$ axis is much larger than that i%n the $xy$ plane
%so that the motion of an electron along the $z$ direction is negligible.
In the case of $p_x$-wave superconductor, we assume for simplicity
that triplet Cooper pairs consist of two electrons with opposite  spin
projections on $z$-axis, i.e. $S_{Z}=0$.
These assumptions do not limit the generality of the discussion
below. We consider a perfect N/S interface with full transmissivity,
while it can be shown that characteristic behavior of Cooper pairs
remains qualitatively unchanged even in the presence of a potential
barrier at the N/S interface.

%The reflection coefficient of the junction for the quasiparticle for
%injection angle $\theta $ is given by $R=Z^{2}/(Z^{2}+4\cos ^{2}\theta )$
%with $Z=2H/v_{F}$, where $\theta $ $(-\pi /2<\theta <\pi /2)$ is measured
%from the normal to the interface and $v_{F}$ is the Fermi velocity.
The quasiclassical Green's functions~\cite{Quasi} in a normal metal (N) and
a superconductor (S) are parameterized as
\begin{equation}
\hat{g}_{\pm }^{(i)}=f_{1\pm }^{(i)} \hat{\tau}_{1}+f_{2\pm }^{(i)}\hat{\tau}%
_{2} +g_{\pm }^{(i)}\hat{\tau}_{3},\ \ (\hat{g}_{\pm }^{(i)})^{2}=\hat{1},
\end{equation}%
where a superscript $i (= N, S)$ refers to N and S, $\hat{\tau}_{j}$ ($j=1-3$)
are the Pauli matrices, and $\hat{1}$ is a unit matrix.
The subscript $+ (-) $ denotes a moving direction of a quasiparticle in
the $x$ direction~\cite{Quasi}, and $\bar{\Delta}_{+}(x)$ ($\bar{\Delta}_{-}(x)$) is
the pair potential for a left (right) going quasiparticle.
In a normal metal, $\bar{\Delta}_{\pm}(x)$ is set to zero because the
pairing interaction is absent there. The Green's
functions can be expressed in terms of the Ricatti parameters~\cite{Ricatti},
\begin{align}
f_{1\pm }^{(i)} &= \mp \nu_i[\Gamma_{\pm }^{(i)}(x)+\zeta_{\pm }^{(i)}(x)]
/[1+\Gamma_{\pm }^{(i)}(x)\zeta_{\pm }^{(i)}(x)], \\
%f_{1\pm}^{(N)}&= \pm[\Gamma_{\pm}^{(N)}(x) + \zeta_{\pm }^{(N)}(x)]
%/[1 + \Gamma_{\pm }^{(N)}(x)\zeta_{\pm }^{(N)}(x)],  \notag \\
f_{2\pm}^{(i)}&=i [\Gamma_{\pm }^{(i)}(x)-\zeta_{\pm }^{(i)}(x)]
/[1 + \Gamma_{\pm }^{(i)}(x)\zeta_{\pm }^{(i)}(x)],  \\
g_{\pm}^{(i)}&=[1 - \Gamma_{\pm }^{(i)}(x)\zeta_{\pm }^{(i)}(x)]
/[1 + \Gamma_{\pm }^{(i)}(x)\zeta_{\pm }^{(i)}(x)],
\end{align}
with $\nu_i=1$ for $i=S$ and $\nu_i=-1$ for $i=N$.
The parameters $\Gamma_{\pm }^{(i)}(x)$ and $\zeta_{\pm }^{(i)}(x)$
obey the Eilenberger equation of the Ricatti type~\cite{Ricatti},
%\begin{align}
%v_{Fx}\partial _{x}\Gamma_{\pm }^{(N)}(x) &
%=2i \varepsilon \Gamma_{\pm }^{(N)}(x)  \label{eq.1a} \\
%v_{Fx}\partial _{x}\zeta_{\pm }^{(N)}(x) &= -2i\varepsilon \zeta_{\pm }^{(N)}(x%),  \label{eq.1b}
%\end{align}
%and in the S region,
%
\begin{align}
i v_{Fx}\partial_{x}\Gamma_{\pm }^{(i)}(x)
& =-\bar{\Delta}_{\pm }(x)
[1+(\Gamma_{\pm}^{(i)}(x))^{2}] +2\varepsilon \nu_i \Gamma_{\pm }^{(i)}
(x), \nonumber
 \\
i v_{Fx}\partial _{x}\zeta_{\pm }^{(i)}(x) &
=-\bar{\Delta}_{\pm }(x)
[1 + (\zeta_{\pm}^{(i)}(x))^{2}] -2\varepsilon \nu_i \zeta_{\pm }^{(i)}(x),
\nonumber
\end{align}
with $\bar{\varepsilon}=\varepsilon+i\delta_0$
where $\varepsilon$ is the energy of a quasiparticle measured from the Fermi level
and
$\delta_0$ is the inverse of the mean free time due to impurity scattering.
In the clean limit, we consider $\delta_0 \ll \Delta_0$.
Boundary condition at $x=-L$ is given by
$\zeta_{\pm}^{(N)}(-L)=-\Gamma_{\mp}^{(N)}(-L)$.
Boundary condition at the N/S interface becomes
$\zeta_{\pm}^{(S)}(0)=-\Gamma_{\pm}^{(N)}(0)$ and
$\zeta_{\pm}^{(N)}(0)=-\Gamma_{\pm}^{(S)}(0)$.
%
%, are
%\begin{equation}
%\zeta_{\pm }^{(S)}(0)
%=-\frac{(1-R)\Gamma_{\pm }^{(N)}(0)+[R -\Gamma_{+}^{(N)}(0)\Gamma_{-}^{(N)}(0)]%
%\Gamma_{\mp}^{(S)}(0)}{%
%[1- R\Gamma_{+}^{(N)}(0)\Gamma_{-}^{(N)}(0)]
%-(1-R)\Gamma_{\mp}^{(N)}(0)\Gamma_{\mp}^{(S)}(0)}
%\end{equation}%
%and
%\begin{equation}
%\zeta_{\pm }^{(N)}(0)
%=-\frac{(1-R)\Gamma_{\pm }^{(S)}(0)+
%[R - \Gamma_{+}^{(S)}(0)\Gamma_{-}^{(S)}(0)]\Gamma_{\mp}^{(N)}(0)}
%{[1 - R\Gamma_{+}^{(S)}(0)\Gamma_{-}^{(S)}(0)]
%- (1-R)\Gamma_{\mp}^{(N)}(0)\Gamma_{\mp }^{(S)}(0)},
%\end{equation}%
%where $R$ is the reflection coefficient at the interface.
The pair potential $\bar{\Delta}_{\pm}(x)$ is expressed by
$\bar{\Delta}_{\pm}(x)=\Delta (x)\Phi _{\pm }(\theta )\Theta(x)$,
where a form factor $\Phi_{\pm }(\theta )$ is given by $\Phi_{\pm
}(\theta )=1$ for $s$-wave symmetry and $\pm \cos \theta $ for
$p_{x}$-wave one with $\theta$ being an incident angle of a
quasiparticle measured from the $x$ direction. Bulk pair potential
is $\Delta(\infty)=\Delta_{0}$, and we determine the spatial
dependence $\Delta (x)$ in a self-consistent way.
\par
For $x \gg  L_0$, the angular structure of $f_{2\pm}^{(i)}$ follows
that of the pair potential, whereas $f_{1\pm}^{(i)}$ is zero with
$L_{0}=v_{F}/2\pi T_{C}$ being a coherence length and $T_{C}$ being
the transition temperature. The pairing function $f_{1\pm }^{(i)}$
is generated by inhomogeneity in a system and thus has a finite
value only near the interface and in a normal metal. Recent
study~\cite{OddNS} showed that $f_{1\pm }^{(i)}$ has an
odd-frequency symmetry.
%$f_{1\pm}^{(i)}(-\varepsilon )=-[f_{1\pm}^{(i)}(\varepsilon )]^{\ast }$,
%$f_{2\pm }^{(i)}(-\varepsilon )=[f_{2\pm}^{(i)}(\varepsilon )]^{\ast }$
%with $f_{1\pm}^{(i)}(-\varepsilon )=f_{1\pm}^{(i)}$, and
%$f_{2\pm }^{(i)}(-\varepsilon )=f_{2\pm}^{(i)}$.
%
Generally speaking, functions $f_{1\pm}^{(i)}$ and $f_{2\pm }^{(i)}$
have opposite parities. If a superconductor has $s$- ($p_{x}$)-wave
symmetry, the induced odd-frequency component has the odd (even)
parity, respectively. Pairing function $f_{1}$ is defined in the
angular domain of $-\pi/2 \le \theta < 3\pi/2$. We denote
$f_{1}(\theta)$ by $f_{1+}(\theta)$ in the angle range $-\pi /2 \le
\theta <\pi /2$ and $f_{1}(\theta)=f_{1-}(\pi-\theta)$ for $ \pi/2
\le \theta <3\pi /2$. The angular structure of functions $f_2$ and
$g$ is defined in the same manner. LDOS is given by the relation
$\rho_{L}(\theta) = {\rm Real}[g(\theta)]$.
%
%
%In the following, we calculate $f_{1}$, $f_{2}$, and $\rho_{L}$.
%self-consistently calculate
%the spatial dependence of the pair potential.
%After that we calculate the spectral properties of pair amplitudes and LDOS.
%
In what follows, we fix temperature $T=0.05T_{C}$, the length of the
normal region $L=5L_{0}$, and $\delta_{0} = 0.01\Delta_{0}$.
\par
\begin{figure}[tb]
\begin{center}
\scalebox{0.8}{
\includegraphics[width=10cm,clip]{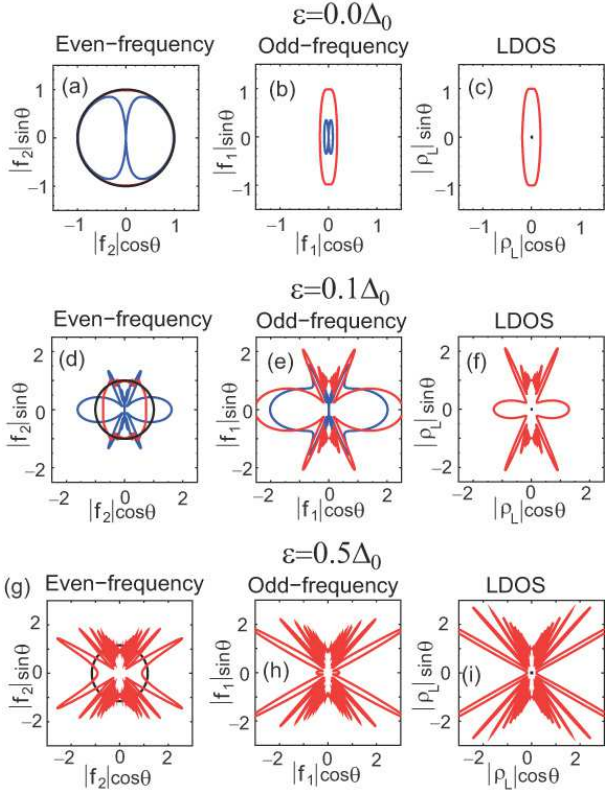}
}
\end{center}
\caption{(Color online)
The results for $s$-wave symmetry.
The shape of the even-frequency pair amplitude [(a),(d),(g)],
the odd-frequency one [(b),(e),(h)], and the
angle resolved local density of state [(c),(f),(i)].
Red lines: $x=0$ (at N/S interface),
blue lines: $x=-L/2$ (in a normal metal) and
black lines: $x=\infty$ (in a superconductor).
$\varepsilon=0$ for (a), (b) and (c),
$\varepsilon=0.1\Delta_{0}$ for (d), (e) and (f), and
$\varepsilon=0.5\Delta_{0}$ for (g), (h) and (i).
The angle $\theta$ is measured from the $x$-axis.
}
\label{fig:2}
\end{figure}
\begin{figure}[tb]
\begin{center}
\scalebox{0.8}{
\includegraphics[width=10cm,clip]{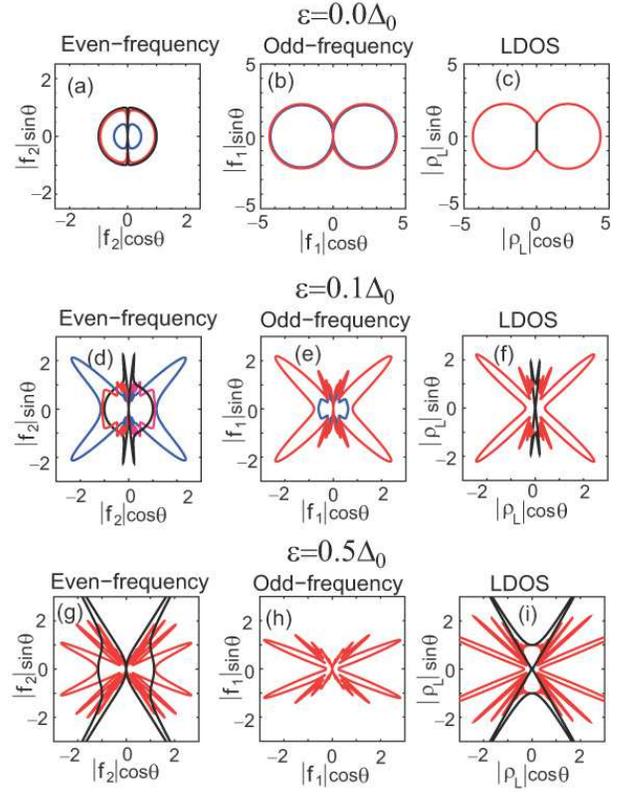}
}
\end{center}
\caption{(Color online)
The results for $p_x$-wave symmetry.
The notations are the same as in Fig.~\ref{fig:2}.
}
\label{fig:3}
\end{figure}

%The boundary condition at $x$=0 can be easily written as
%$\zeta_{\pm}^{(S)}(-L)=-\Gamma_{\pm}^{(N)}(-L)$,
%$\zeta_{\pm}^{(S)}(-L)=-\Gamma_{\pm}^{(N)}(-L)$. \par
In  Figs.~\ref{fig:2} and ~\ref{fig:3} we show polar plots of
$f_{1}$, $f_{2}$, and $\rho_{L}$ in a $s$-wave and $p_{x}$-wave
junctions for several choices of $\varepsilon$ and $x$. Black, red
and blue lines represent, respectively, the results for $x=\infty$
(superconductor), $x=0$ (interface), and $x=-L/2$ (normal metal).
The odd-frequency component is always absent for $x=\infty$. Since
$\rho_{L}$ is independent of $x$ in N, the resulting value of
$\rho_{L}$ at $x=0$ is equal to that  at $x=-L/2$.

\par First, we
focus on the $s$-wave case (Fig.~\ref{fig:2}). At $\varepsilon=0$ in
Fig.~\ref{fig:2}(a), the even-frequency component $f_2$ has a
circular shape reflecting $s$-wave symmetry for $x=\infty$ and  0.
However, at $x=-L/2$, the shape is no longer a simple circular but a
double distorted circles. The shape of $f_2$ in superconductor
always has the circular shape independent of $\varepsilon$ as shown
with black broken lines in Fig.2(a), (d), and (g). At
$\varepsilon=0.1\Delta_0$ in Fig.2(d), $f_2$ at the interface (red
line) slightly deviates from the circular shape, while the shape in
N drastically changes. The tendency is more remarkable at
$\varepsilon=0.5\Delta_0$ in Fig.2(g). The butterfly-like pattern
seen in $f_2$ at the interface (red line in Fig.2(g)) is completely
different from the original circular shape in a superconductor. At
$\varepsilon=0$, $f_{1}$ at the interface becomes ellipsoidal as
shown in a red line in Fig.2(b). The shape of $f_{1}$ at $x=0$ and
$x=-L/2$ shows the butterfly-like pattern (Fig. 2(e)). For
$\varepsilon=0.5\Delta_{0}$, the line shape of $f_{1}$ has many
spikes as shown in  Fig. 2(h). Such anisotropic property of $f_{1}$
and $f_2$ affects the LDOS as shown in Fig.2(c), (f), and (i). In
particular, LDOS at the interface for $\varepsilon=0.5\Delta_0$ (red
line in (i)) strongly deviates from the circular shape. The LDOS in
a superconductor vanishes at $\varepsilon <\Delta_0$ (black lines in
Fig.2(f) and (i)). At the interface, the shape of $\rho_{L}$ in
Fig.2(c) is quite similar to that of $f_{1}$ in Fig.2(b).
%For $\varepsilon=0.1\Delta_{0}$,
%shapes of $f_{2}$ are not so much changed seriously(blue and  line in (d)).
%Only at the interface, several spikes appear (red line).
%The corresponding odd-frequency component is  much enhanced
%at the interface (red line in (e)).
%The corresponding $\rho_{L}$ has a rather similar
%shape (red line in (c)).
%For $\varepsilon=0.5\Delta_{0}$,
%shapes of $f_{2}$ near the interface
%become very complex.
%They are far from the circular.
%Also the corresponding $f_{1}$ and $\rho_{L}$
%have very complex line shapes
%including many spikes (red curves in (f),(g) and (h)).
%
These profiles can be qualitatively understood as follows. At $x=0$,
the following relations hold
%\begin{equation}
%f_{1\pm}^{(N)}=\pm \Gamma
%\frac{1 - \alpha^{2} }{1 - \Gamma^{2}\alpha^{2}},  \
%f_{2\pm}^{(N)}= i \Gamma
%\frac{1 + \alpha^{2} }{1 - \Gamma^{2}\alpha^{2}},  \
%\end{equation}
%\begin{equation}
%g_{\pm}=
%\frac{1 + \alpha^{2}\Gamma^{2} }
%{1 - \Gamma^{2}\alpha^{2}},
%\label{eq.g}
%\end{equation}
\begin{equation}
f_{1\pm}^{(N)}=\pm \Gamma \frac{1 - \alpha^{2} }{\Xi},\
f_{2\pm}^{(N)}= i \Gamma
\frac{1 + \alpha^{2} }{\Xi},  \
g_{\pm}^N=
\frac{1 + \alpha^{2}\Gamma^{2} }
{\Xi},
\label{eq.g}
\end{equation}
with $\Xi=1 - \Gamma^{2}\alpha^{2}$, $\Gamma=\Gamma_{\pm}^{(S)}(0)$
and $\alpha=\exp[2i \varepsilon L/(v_{F}\cos\theta)]$. For
$\varepsilon \ll \Delta_{0}$, $\Gamma \sim 1/i$ and $f_{1\pm} \sim i
g_{\pm}$ are satisfied. Thus shape of function $f_{1}$ is similar to
that of $\rho_{L}$. This argument seems to be valid even for
$\varepsilon=0.1\Delta_0$ in Fig.2(e) and (f), and for
$\varepsilon=0.5\Delta_0$ in Fig.2(h) and (i). The oscillating
behavior in $f_1$, $f_2$, and $\rho_L$ is more remarkable at
$\varepsilon =0.5\Delta_{0}$. Although we do not present calculated
results of $f_{1}$ and $f_{2}$ at $x=-L/2$ for
$\varepsilon=0.5\Delta_0$, the butterfly-like pattern with many
spikes in the pairing functions can be seen also in a normal metal.
The directions of the spin projections in LDOS are characterized by
small value of $\Xi$, which has close relation to the formation of
the bound states~\cite{Rowell}. For $\theta \sim \pm \pi/2$,
$\alpha$ oscillates rapidly with small variation of $\theta$, which
explains the fine structures in LDOS around $\theta=\pm\pi/2$. The
quasiparticle interference effect is a source of bound state
formation in a normal metal. As a result, the circular shape of
Cooper pairs in $s$-wave superconductor is modified into the
butterfly-like pattern in a normal metal.
\par
Next, we discuss the results for $p_{x}$-wave junctions shown in
Fig.~\ref{fig:3}. In a superconductor ($x=\infty$), functions
$f_{2}$ and $g$ are given by $\Delta_{0} \cos\theta /
\sqrt{\varepsilon^{2} - \Delta_{0}^{2}\cos^{2}\theta}$ and
$\varepsilon / \sqrt{\varepsilon^{2} -
\Delta_{0}^{2}\cos^{2}\theta}$, respectively. As shown by black
lines in Fig.3(d), (f), (g) and (i), the amplitudes of $f_{2}$ and
$g$ become large along the directions $\theta
=\cos^{-1}(\varepsilon/\Delta_0)$. At $\varepsilon=0$ and $x=0$,
formation of a mid-gap Andreev resonant state~\cite{TK95,ZES}
significantly enhances the amplitudes of $f_{1}$ and $\rho_{L}$
compared to that of $f_{2}$ as shown by red lines in Fig.3(a), (b),
and (c). For $\varepsilon=0$ and $\varepsilon=0.1\Delta_0$, the
shapes of $f_{1}$ and $\rho_{L}$ at the N/S interface are similar to
those in $s$-wave superconductor junctions (red lines in Fig.3(b),
(c), (e) and (f)]. The shape of $f_{1}$ and $f_{2}$ in the N region
is rather complex too. At $\varepsilon=0.5\Delta_{0}$, $f_{1}$,
$f_{2}$ and $\rho_{L}$ also exhibit the butterfly-like patterns as
shown by red lines in Fig.3(g), (h), and (i). Similar to $s$-wave
case, functions $f_{1}$ and $f_{2}$ in N have a complex line shapes
with many spikes.
%%%%%%%%%%%%%%%%%%%%%%%%%%%%%%%%%%%%%%%%%%%%%%%%%%%%%%%%%%
%CHANGE
%%%%%%%%%%%%%%%%%%%%%%%%%%%%%%%%%%%%%%%%%%%%%%%%%%%%%%%%
%------------------------------
The butterfly-like pattern in angle resolved local density of states
could be observed through the LDOS in magnetic fields.

\par Here, we propose an experimental setup of scanning tunneling
spectroscopy (STS) in the presence of the magnetic field. As shown
in Fig.~\ref{fig:1}, magnetic field is applied parallel to the N/S
plane. Tunneling current at a fixed bias voltage is measured as a
function of the angle $\phi$ between the $x$ axis and the direction
of magnetic field. The vector potential in this configuration is
given by $(A_{x},A_{y})=-\lambda H \exp(-z/\lambda)(\sin \phi, \cos
\phi)$ \cite{Doppler,Magneto}. We assume that thickness of a quasi
two-dimensional superconductor is sufficiently small compared to the
magnetic field penetration depth $\lambda$. Magnetic field shifts
the quasiparticle energy $\varepsilon$ to $\varepsilon - H
\Delta_{0} \sin(\phi-\theta)/B_{0}$ where
 $B_{0}= h/(2e\pi^{2}\xi\lambda)$ and $\xi=\hbar v_F/\pi \Delta_0$.
For typical values of $\xi \sim \lambda \sim 100$nm,
the magnitude of $B_{0}$ is of the order of 0.02Tesla.
%
%
%Concerning the estimate of magnetic field, let's take \xi ~ \lambda ~ 100 nm ~ %10^{-7} m, then
%
%we get for the field  \Fhi_0/ pi^2 \xi \lambda ~ 2 10^{-15}/ pi^2 10^{-14} ~ 0.%02 Tesla,%
%
%or just by order of magnitude 10^{-2} Tesla
Local density of state observed in STS experiments is given by,
$ \rho(\phi)= \int^{3\pi/2}_{-\pi/2} \rho_{L}(\theta,\phi) d\theta$. \par
\begin{figure}[tb]
\begin{center}
\scalebox{0.8}{
\includegraphics[width=9cm,clip]{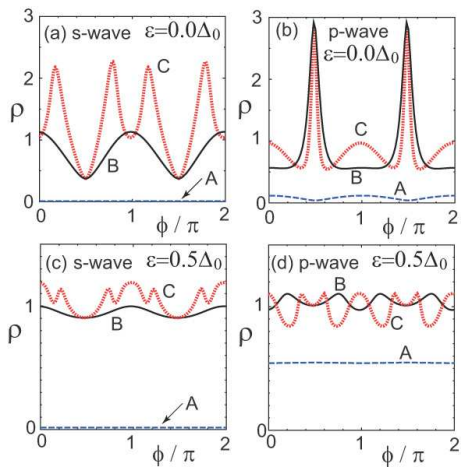}
}
\end{center}
\caption{(Color online)
Magneto-tunneling conductance as a function of the orientation angle
of magnetic field $\phi$ in a spin-singlet $s$-wave junction [(a) and (c)] and
in a spin-triplet $p_{x}$-wave one [(b) and (d)].
$\varepsilon=0$ in the cases (a) (b) and $\varepsilon=0.5\Delta_{0}$ in the cases (c), (d).
A: $x=\infty$ (in a superconductor) for $H=0.1B_{0}$,
B: $x=0$ (at N/S interface) for $H=0.05B_{0}$, and
C: $x=0$ for $H=0.1B_{0}$.
}
\label{fig:4}
\end{figure}
%-------------------
%-------------------
In Fig.~\ref{fig:4}, $\rho$ is plotted as a function of $\phi$. For
$\varepsilon=0$, $\Delta_{0} H/B_{0}$ gives the effective energy of a
quasiparticle. In an $s$-wave superconductor, it is evident that
$\rho$ is independent of $\phi$ and the amplitude of LDOS is
vanishingly small, see curve A in Fig.~\ref{fig:4}(a). In a
$p_{x}$-wave superconductor, as shown by curve A in
Fig.~\ref{fig:4}(b), $\rho$ depends slightly on $\phi$ due to
quasiparticle excitations near the nodal points at the Fermi surface
~\cite{Vekhter}.
%
%
%The results of line A in Fig.~\ref{fig:4}(b) becomes minima
%when $H$ is parallel to the nodal directions ($\theta=\pm \pi/2$)
At the N/S interface, $\rho$ in the $s$-wave case has minima at
$\phi=\pi/2$ and $3 \pi/2$ and maxima at $\phi=0$ and $\pi$, as shown
by curve B in Fig.~\ref{fig:4}(a). On the other hand, in the
$p_{x}$-wave case, $\rho$ has maxima at $\phi=\pi/2$ and $3 \pi/2$,
see curve B in Fig.~\ref{fig:4}(b). This behavior can be explained
in the following way. Eq.~(\ref{eq.g}) yields approximate expression
for $\rho$ at $H \ll B_{0}$
\begin{equation}
\rho \sim \int^{\pi/2}_{-\pi/2} \frac{ 2(1 - a^{2})\; d\theta}
{1 + a^{2} \pm 2a \cos [C \sin(\theta-\phi)/\cos \theta]},
\end{equation}
where $a=\exp(-4L \delta_{0}/\hbar |{v_{Fx}}| )$, $C=4LH/\hbar
|v_{Fx} |$, and the sign in the denominator is $+$ ($-$) for
$s$-wave ($p_{x}$-wave) junctions. We note that $a$ is a positive
number almost independent of $\theta$ and $C$ is a small
positive number. For $\phi=n\pi + \pi/2$ with integer $n$, the magnitude of the
argument of cosine function in the denominator becomes small and
the denominator is reduced to $(1 \pm a)^{2}$ for $s$-wave ($p_{x}$-wave)
case. When  $\phi$ deviates from $n\pi + \pi/2$, the cosine function
decreases and therefore $\rho$ has a dip (peak) at $\phi=\pi/2$ and
$3\pi/2$ for $s$-wave ($p_{x}$-wave) cases as shown by curve B in
Fig.~\ref{fig:4}(a) ((b)). Similar argument also explains the
maximum (minimum) of $\rho$ at $\phi=n\pi$ in $s$-wave
($p_{x}$-wave) junctions.
%In both $s$- and $p_x$-wave cases, $\rho$ takes maximum
%where magnetic field is applied parallel to the direction
%in which $\rho_{L}$ has a maximum at $\varepsilon=0$.
When the field $H$ is increases further (curves C in Figs.4(a) and
(b)), the shape of the $\rho$ exhibits more complex behavior
reflecting the butterfly-like pattern of $\rho_{L}$ shown in
Figs.~\ref{fig:2} and \ref{fig:3}, because the increase of $H$ has
qualitatively similar effect as the increase of $\varepsilon$.
\par

The complicated oscillating features are seen much more clear at
$\varepsilon=0.5\Delta_{0}$. As shown by curves $C$ in Figs. 4(c)
and (d), the period of oscillations becomes shorter than that in
Figs.~\ref{fig:4}(a) and (b), reflecting the butterfly-like patterns
in Figs.~\ref{fig:2}(i) and Figs.~\ref{fig:3}(i). At the same time,
the magnitude of the oscillations at $\varepsilon=0.5\Delta_{0}$
becomes smaller than that at $\varepsilon=0$, since the integration
with respect to $\theta$ averages the butterfly-like pattern in
$\rho_L$. The discussed features in LDOS at the N/S interface differ
strongly from those in a bulk superconductor. We conclude that STS
experiments in magnetic field should resolve the remarkable
deformation of Cooper pairs. \par

In summary, we have studied the Cooper pair shape in
normal-metal/superconductor (N/S) junctions by using the
quasiclassical Green's function formalism. The quasiparticle
interference leads to striking deformations in the shape of a Cooper
pair wave function in a normal metal. As a consequence, the angle
resolved local density of states exhibits the butterfly-like
patterns. We also show that the anisotropic shape of Cooper pairs
could be resolved by scanning tunneling spectroscopy experiments in
magnetic field. The Cooper pair deformation is a common feature of
non-uniform superconducting systems in the clean limit. This
provides a key concept to explore new quantum interference phenomena
in superconducting nanostructures.

%---------------------


\begin{thebibliography}{99}
%\bibitem{} % General discussion of the Cooper pair
%%%%%%%%%%%%%%%%%%%%%%%%%%%%%%%%%%%%%%%%

%\bibitem{Tinkham}
%M. Tinkham, {\it Introduction to Superconductivity} (2nd edn),
%McGraw-Hill, New York (1996).

\bibitem{deGennes}
de Gennes
{\it Superconductivity of Metal and Alloys},
W.A. Benjamin, Inc. (1966).


\bibitem{Sigrist} M. Sigrist and K. Ueda, Rev. Mod. Phys. \textbf{63} 239
(1991).

%%%%%%%%%%%%%%%%%%%%%%%%%%%%%%%%%%%%%%%%%%
% Odd freqauency superconductor
%%%%%%%%%%%%%%%%%%%%%%%%%%%%%%%%%%%%%%%%%%%%%
\bibitem{Berezinskii} V. L. Berezinskii, JETP Lett. \textbf{20}, 287 (1974).

\bibitem{Bergeret}
F. S. Bergeret, A. F. Volkov, and K. B. Efetov, Rev. Mod. Phys.
\textbf{77} 1321 (2005).

\bibitem{Golubov2007} Y. Tanaka and A.A. Golubov, Phys. Rev. Lett. \textbf{98}%
, 037003 (2007); Y. Tanaka, A.A. Golubov, S. Kashiwaya and M. Ueda,
Phys. Rev. Lett. {\textbf 99} 037005 (2007);
M. Eschrig, T. Lofwander, Th. Champel, J.C. Cuevas and G. Schon,
J. Low Temp. Phys. \textbf{147} 457 (2007).

\bibitem{Belzig}
W. Belzig, C. Bruder, and A. L. Fauchere, Phys. Rev. B \textbf{58}
14531 (1998).

\bibitem{GK} A.A.Golubov and M.Yu.Kupriyanov, JETP Lett \textbf{67}, 501
(1998).

\bibitem{Quasi}J.W. Serene and D. Rainer, Phys. Rep. \textbf{101} 221
(1983).

\bibitem{Ricatti} M. Eschrig, Phys. Rev. B \textbf{61} 9061 (2000),
A.Shelankov and M. Ozana, Phys. Rev. B \textbf{61}, 7077 (2000); N.
Schopohl and K. Maki, Phys. Rev. B \textbf{52}, 490 (1995).

\bibitem{OddNS}
Y. Tanaka, Y. Tanuma and A.A. Golubov, Phys. Rev. B {\textbf 76}
054522 (2007).

%\bibitem{McMillan} W. L. McMillan, Phys. Rev. \textbf{175}, 537 (1968).

\bibitem{Rowell} J. M. Rowell and W. L. McMillan, Phys. Rev. Lett. \textbf{16%
}, 453 (1966); J. M. Rowell, Phys. Rev. Lett. \textbf{30}, 167 (1973).
%%%%%%%%%%%%%%%%%%%%%%%%%%%%%%%%%%%%%%%%%%%%%%%%%%%%%%%%
% MARS
%%%%%%%%%%%%%%%%%%%%%%%%%%%%%%%%%%%%%%%%%%%%%%%%%%%%%%%%

\bibitem{TK95} Y. Tanaka and S. Kashiwaya, Phys. Rev. Lett. \textbf{74},
3451 (1995); S. Kashiwaya and Y. Tanaka, Rep. Prog. Phys. \textbf{63}, 1641
(2000), T. L\"{o}fwander, V. S. Shumeiko, and G. Wendin, Supercond. Sci.
Technol. \textbf{14}, R53 (2001).

\bibitem{ZES} L.J. Buchholtz and G. Zwicknagl, Phys. Rev. B \textbf{23},
5788 (1981); J. Hara and K. Nagai, Prog. Theor. Phys. \textbf{74}, 1237
(1986); C.R. Hu, Phys. Rev. Lett. \textbf{72}, 1526 (1994); C. Bruder, Phys.
Rev. B \textbf{41}, 4017 (1990).

%%%%%%%%%%%%%%%%%%%%%%%%%%%%%%%%%%%%%%%%%%%%%%%%%%%%%%%%
% Experiment of ZEP
%%%%%%%%%%%%%%%%%%%%%%%%%%%%%%%%%%%%%%%%%%%%%%%%%%%%%%%%
%; Yu.S. Barash, A.A. Svidzinsky, and H.
%Burkhardt, Phys. Rev. B \textbf{55}, 15282 (1997).

%%%%%%%%%%%%%%%%%%%%%%%%%%%%%%%%%%%%%%%%%%%%%%%

% C. Iniotakis, S. Graser,
%T. Dahm, and N. Schopohl Phys. Rev. B \textbf{71}, 214508 (2005).

%%%%%%%%%%%%%%%%%%%%%%%%%%%%%%%%%%%%%%%%%%%%%%%
%
%\bibitem{Eilen} G. Eilenberger, Z. Phys. \textbf{214}, 195 (1968).
%
%%%%%%%%%%%%%%%%%%%%%%%%%%%%%%%%%%%%%%%%%%%%%%%%%%%%%%%%%%
% Proximity
%%%%%%%%%%%%%%%%%%%%%%%%%%%%%%%%%%%%%%%%%%%%%%%%%%%%%%%

%\bibitem{proximityd} Y. Tanaka, Y.V. Nazarov and S. Kashiwaya, Phys. Rev.
%Lett. \textbf{90}, 167003 (2003); Y. Tanaka, Y.V. Nazarov, A.A. Golubov, and
%S. Kashiwaya, Phys. Rev. B \textbf{69}, 144519 (2004).

\bibitem{Doppler}
M. Fogelstr\"{o}m, D. Rainer and J. A. Sauls, Phys. Rev. Lett. 79 281
(1997).

\bibitem{Magneto}
Y. Tanaka, Y. Tanuma, K. Kuroki and S. Kashiwaya,
J. Phys. Soc. Jpn. \textbf{71}, 2102 (2002).
%Y. Tanuma, K. Kuroki, Y. Tanaka, R. Arita, S. Kashiwaya and H. Aoki,
%Phys. Rev. B \textbf{66} (2002) 094507.
%Y. Tanuma, Y. Tanaka, K. Kuroki and S. Kashiwaya, Phys. Rev. B \textbf{66} (200%2) 174502.

\bibitem{Vekhter}
I. Vekhter, P. J. Hirschfeld, J. P. Carbotte, and E. J. Nicol,
Phys. Rev. B 59 (1999) R9023.

\end{thebibliography}
\end{document}